\documentstyle[12pt]{article}

\newcommand{\EQ}{\begin{equation}}
\newcommand{\sh}{\sinh}
\newcommand{\th}{\tanh}
\newcommand{\ch}{\cosh}
\newcommand{\EN}{\end{equation}}

\newcommand{\bear}{\begin{eqnarray}}
\newcommand{\ear}{\end{eqnarray}}

\begin{document}

\topmargin 0pt
\oddsidemargin 5mm
\newcommand{\NP}[1]{Nucl.\ Phys.\ {\bf #1}}
\newcommand{\PL}[1]{Phys.\ Lett.\ {\bf #1}}
\newcommand{\NC}[1]{Nuovo Cimento {\bf #1}}
\newcommand{\CMP}[1]{Comm.\ Math.\ Phys.\ {\bf #1}}
\newcommand{\PR}[1]{Phys.\ Rev.\ {\bf #1}}
\newcommand{\PRL}[1]{Phys.\ Rev.\ Lett.\ {\bf #1}}
\newcommand{\MPL}[1]{Mod.\ Phys.\ Lett.\ {\bf #1}}
\newcommand{\JETP}[1]{Sov.\ Phys.\ JETP {\bf #1}}
\newcommand{\TMP}[1]{Teor.\ Mat.\ Fiz.\ {\bf #1}}
     
\renewcommand{\thefootnote}{\fnsymbol{footnote}}
     
\newpage
\setcounter{page}{0}
\begin{titlepage}     
\begin{flushright}
ITFA-97-15
\end{flushright}
\vspace{0.5cm}
\begin{center}
\large{  On the solution of a supersymmetric model of correlated electrons } \\
\vspace{1cm}
\vspace{1cm}
 {\large M. J.  Martins$^{1,2}$  and P.B. Ramos$^{2}$} \\
\vspace{1cm}
\centerline{\em ${}^{1}$ Instituut voor Theoretische Fysica, Universiteit van Amsterdam  }
\centerline{\em  Valcknierstraat 65, 1018 XE Amsterdam, The Netherlands}
\centerline{ \em and }
\centerline{\em ${}^{2}$ Departamento de F\'isica, Universidade Federal de S\~ao Carlos}
\centerline{\em Caixa Postal 676, 13565-905, S\~ao Carlos, Brasil}
\vspace{1.2cm}   
\end{center} 
\begin{abstract}
We consider the exact solution of a model of correlated electrons based on
the superalgebra $Osp(2|2)$. The corresponding Bethe ansatz
equations have an interesting form. We 
derive an expression for the ground state energy  at half filling. We also
present the eigenvalue of the transfer matrix commuting with the Hamiltonian.
\end{abstract}
\vspace{.2cm}
\centerline{PACS numbers: 71.27.+a, 75.10.Jm}
\vspace{.2cm}
\centerline{March 1997}
\end{titlepage}

\renewcommand{\thefootnote}{\arabic{footnote}}
Recently supersymmetric generalizations of the Hubbard model have attracted
considerable interest for their possible relevance in describing correlated
electron systems. A successful example is the Essler, Korepin and Schoutens
\cite{EKS} extended Hubbard model due to its superconductive properties. This
model is based on the $Sl(2|2)$ superalgebra and is exactly solvable in one
dimension by the Bethe ansatz \cite{EKS,EKS1}. Other interesting system is a
supersymmetric free-parameter model based on the continuous representation
of the $gl(2|1)$ superalgebra discussed by Bracken et al \cite{BRA}. We recall
that, for a special value of the coupling constant, this model was also considered by Karnaukhov \cite{KA}. Shortly afterwords, an anisotropic deformation of
this model was found \cite{KUU,ZU1} as well as its solution by the coordinate
Bethe ansatz approach \cite{KUU,FRA}.

The main feature of these models is that in one dimension they are integrable
and therefore they provide us with non-perturbative information concerning
physical properties. In fact, they can be derived from 
supersymmetric solutions of the Yang-Baxter 
equation invariant either by the $Sl(2|2)$ or by the
$Osp(2|2)$ superalgebras \cite{BRA,ZU3,MA,DEG}. However, at least from the Bethe
ansatz point of view, there exists a supersymmetric solution which has not
yet been discussed in the literature so far. This solution was found sometime
ago by Deguchi et al \cite{DEG} in the context of a $q$-deformed $Osp(2|2)$
superalgebra and very recently has been interpreted as being invariant by the
twisted $U_q[Sl(2|2)^{(2)}]$ superalgebra \cite{ZU4}. It seems that
this model exhausts the cases which can be derived by exploiting the $Osp(2|2)$
superalgebra \cite{MA,ZU4}. This then provides us with an extra motivation
for studying this supersymmetric system. The purpose of this paper is to
discuss this solution in the context of an exactly solvable model of correlated
electrons by presenting its Bethe ansatz solution.

The one dimensional lattice Hamiltonian is derived from the logarithmic 
derivative of the $R$-matrix solving the supersymmetric Yang-Baxter equation.
For  explicit expressions we 
refer to refs. \cite{DEG,MA,ZU4}. Its turns out that the corresponding Hamiltonian can be written in terms of fermionic operators $c_{i,\sigma}^{\dagger}$
and $c_{i,\sigma}$ acting on the sites $i$ of a chain of length $L$ and
carring spin index $\sigma=\pm$. We found that the Hamiltonian reads as
\bear
H = \sum_{i=1}^{L} \sum_{\sigma=\pm} \left [ c_{i, \sigma}^{\dagger} 
c_{i+1, \sigma} +
h.c  \right ] \left [ 1- n_{i, -\sigma}(1+\sigma V_1)
-n_{i+1, -\sigma}(1-\sigma V_1) \right ] +
\nonumber \\
V_2 \sum_{i=1}^{L} \left [ c_{i,+}^{\dagger} c_{i,-}^{\dagger} 
c_{i+1,-} c_{i+1,+}
- c_{i,+}^{\dagger} c_{i+1,-}^{\dagger} c_{i+1,+} c_{i,-} +h.c \right ] +
\nonumber \\
V_2 \sum_{i=1}^{L} \left [ n_{i,+} n_{i,-} +n_{i+1,+} n_{i+1,-} + n_{i,+} n_{i+1,-} +n_{i,-}n_{i+1,+} -n_{i} -n_{i+1} +1 \right ]
\ear
where $n_{i,\sigma}= c_{i,\sigma}^{\dagger} c_{i,\sigma} $ is the number
operator for electrons with spin $\sigma$ on site $i$ and we write $n_i=
n_{i,+} +n_{i,-}$. This Hamiltonian conserves the number $N_+$ and $N_{-}$ of
electrons with spin up and down, respectively. Later on we will use the
total number of electrons $N_e=N_+ + N_{-}$ and the number $N_{+}$ as the
good ``quantum'' numbers for charactering the spectrum of
the Hamiltonian. We
observe that in the derivation of expression (1) we have made use of convenient
canonical transformations and also we have disconsidered terms ( such as
the one proportional to $n_i -n_{i+1} $ ) which are automatically canceled out
when periodic boundary conditions are imposed.

The Hamiltonian (1) presents extra fine tuned hopping terms, on site
and off site Coulomb interactions. We recall that such interactions resemble
much those appearing in the model of hole superconductivity
proposed by Hirsch \cite{HI}, and they can be derived from first
principles. This model becomes integrable when the couplings $V_1$ and $V_2$
are constrained on the unitary circle, i.e. $V_1^2 +V_2^2$=1. The 
parametrization of these couplings in terms of the $q$-deformed parameter
$q=\exp(i\gamma)$ ($0 \leq \gamma \leq \pi $) of the $Osp(2|2)$ superalgebra
can be read as
\EQ
V_1=  \sin(\gamma), V_2 = \cos(\gamma)
\EN

From equation (1) we see that when $\gamma \rightarrow 0$, this Hamiltonian
reduces to the model proposed by Essler, Korepin and Shoutens \cite{EKS}. This
has been earlier noted in ref. \cite{ZU4}. It  
is still possible to reduce the interval for the anisotropy
$\gamma$ if we perform particle-hole transformations. Indeed, the 
combined set of canonical transformations
(I) $c_{i,\sigma}^{\dagger} \leftrightarrow c_{i,
\sigma}$ and (II) 
 $c_{i,\sigma}^{\dagger} \leftrightarrow (-1)^{i} c_{i,\sigma}$ leads to
conclude  that
the spectrum at certain values of $\gamma$ and $\pi -\gamma$ are related to
each other. More precisely, we have the following identity
\EQ
H(\gamma)= -H(\pi -\gamma) 
\EN
and consequently we can restrict our 
analysis only for the regime $0 \leq \gamma 
\leq \pi/2 $. This identity indicates that the symmetric point $\gamma=\pi/2$
is somewhat special. In fact at $\gamma=\pi/2$ we are left with ``almost''
a free-fermion theory. In this case the spectrum splits in two sectors,
depending on the parity of the total number of electrons in the lattice.
We have two decoupled $XY$ models, with antiperiodic or periodic 
boundary conditions for
$N_e$ even or odd, respectively. We note that $H(\pi/2)$  corresponds also
to the limit $U \rightarrow \infty $ in the Bariev chain \cite{BA}, which is
a special case of a generalized $XY$ model proposed 
long ago by Suzuki \cite{SU}. This means that the Deguchi et al's 
$R$-matrix provides also  embedding (`` covering'' vertex model )
for this peculiar limit of the Bariev chain \cite{SHI}.

We now turn to the diagonalization of Hamiltonian (1) by the coordinate Bethe
ansatz formalism. In other words, we would like to solve the eigenvalue 
problem, $H \Psi =E(L) \Psi $,  provided the wave function in the sector of
$N_e$ electrons distributed on the positions $ 1 \leq x_{Q_1} \leq
x_{Q_2} \leq \cdots \leq x_{Q_{N_e}} \leq L$ has the following form
\EQ
\Psi_{\sigma_1, \cdots, \sigma_{N_e}}(x_{Q_1}, \cdots, x_{Q_{N_e}}) = 
\sum_{P} sgn(P) 
\prod_{j=1}^{N_e} \exp[ip_{P_j} x_{Q_j}] 
A(P|Q)_{\sigma_1, \cdots, \sigma_{N_e}} 
\EN
where  the $P$ summation extends over all the permutations of the
momenta ($P_1, \cdots, P_{N_e}$) and $sgn$ is the sign of the permutation.
When electrons are enough far apart ($ |x_{Q_i} -x_{Q_j}| \geq 2$) it is direct
to derive that the eigenvalues are given by
\EQ
E(L) = \sum_{j=1}^{N_e} 2 \cos(p_j) + V_2 (L-2N_e)
\EN

The next step is to consider the matching condition for the Bethe ansatz wave
function. Since the Hamiltonian (1) has been derived from a 
factorizable $R$-matrix, we can restrict ourselves to the discussion of the
two-body problem (two electrons in the chain). This is given in terms of
the two-body $S$-matrix, which connects the scattering amplitudes
between states $\{ (p_1,\sigma_1); (p_2,\sigma_2) \} $ and 
$\{ (p_2,\sigma_2^{'}); (p_1,\sigma_1^{'}) \} $. We found that the non-null
two-body $S$-matrix elements are given by 
\EQ
S_{++}^{++}(\lambda)=S_{--}^{--}(\lambda)=1,~ 
S_{+-}^{+-}(\lambda)=S_{-+}^{-+}(\lambda)=
\frac{\sh(\lambda)}{\sh(\lambda+2i\gamma)},~
S_{+-}^{-+}(\lambda)=S_{-+}^{+-}(\lambda)=
\frac{\sh(2i \gamma)}{\sh(\lambda+2i\gamma)}
\EN
where $\lambda=\lambda_1 -\lambda_2$ and $\lambda_j$ are the ``dressed''
momenta rapidities which are related to the momenta $p_j$ by the following
relation
\EQ
\exp[ip_j] = \frac{\sh(\lambda_j/2 -i\gamma/2)}{\sh(\lambda_j/2 +i\gamma/2)}
\EN

Now we have the basic ingredients to derive the Bethe ansatz equations.
The next step is to face the problem of diagonalizing the spins degrees of
freedom which are encoded in the $S$-matrix (6). However, since this $S$-matrix
is of $6$-vertex type, this later problem can be solved by standard algebraic
methods \cite{FA}. In the course of the solution we have to introduce new
spin rapidities $\mu_j, j=1, \cdots, N_+$. It turns out 
that the dressed momenta and the spin variables 
satisfy the following nested Bethe ansatz equations
\EQ
\left [ \frac{\sh(\lambda_j/2-i\gamma/2)}{\sh(\lambda_j/2 +i\gamma/2)} 
\right ]^{L} = \prod_{k=1}^{N_{+}} \frac{ \sh(\lambda_j - \mu_k -i\gamma)}
{ \sh(\lambda_j - \mu_k +i\gamma)},~~ j=1, \cdots, N_{e}
\EN
\EQ
\prod_{k=1}^{N_{e}} \frac{ \sh(\mu_j -\lambda_k -i \gamma)}
{\sh(\mu_j -\lambda_k + i \gamma)} = - \prod_{k=1}^{N_{+}} 
\frac{\sh(\mu_j -\mu_k- 2i \gamma)}
{\sh(\mu_j -\mu_k+ 2i \gamma)},~~ j=1, \cdots, N_{+}
\EN
and, in terms of the rapidities $\lambda_j$, the eigenvalues $E(L)$ are given
by
\EQ
E(L) =  \sum_{i=1}^{N_e} \frac{2 \sin^2(\gamma)}{\cos(\gamma) -\ch(\lambda_i)}
+ \cos(\gamma) L
\EN

The Bethe ansatz equations (8,9) have 
the unusual peculiarity that the ``dressed''
momenta variable $\lambda_j$ enters in different ways in 
the momenta (as $\lambda_j/2$) and in the bare 
phase shift (right hand side of (8) ). This is a special
feature which distinguishes this model from the  free-parameter supersymmetric
correlated electron system 
\cite{BRA,KUU,FRA}.
Furthermore, near half-filling we
have determined that while $\lambda_j$ are real roots, the spin variables
$\mu_j$ form strings of the following type
\EQ
\mu_j = \mu_j +i \frac{\pi}{2  }
\EN

By substituting this ansatz in equations (8,9), taking their logarithmic and
afterwards performing the thermodynamic limit $ L \rightarrow \infty $, we
are able to obtain the integral equations for the densities $\sigma(\lambda)$
and $\rho(\mu)$ for the variables $\lambda_j$ and $\mu_j$, respectively. They
are given by
\EQ
\sigma(\lambda) =\frac{\psi_1^{'}(\lambda/2,\gamma/2)}{2 \pi} +
\int_{-\infty}^{+ \infty} d \mu \psi_2^{'}(\lambda-\mu,\gamma) \rho(\mu)
\EN
\EQ
2 \pi \rho(\mu) = \int_{-\infty}^{+\infty} d \lambda \psi_1^{'}(\mu-\lambda,2 \gamma) 
\rho(\lambda) 
+ \int_{-\infty}^{+\infty} d \lambda \psi_2^{'}(\mu-\lambda,\gamma) 
\sigma(\lambda) 
\EN
where $\psi_1(x,\gamma)= 2 \arctan[\cot(\gamma) \tanh(x)] $ 
and $\psi_2(x,\gamma)= 2 \arctan[\tan(\gamma) \tanh(x) ] $ and the symbol $'$
stands for the derivative 
$\psi_{1,2}^{'}(x/a,\gamma)= \frac{d \psi_{1,2}(x/a,\gamma)}{d x}$. These
coupled integral equations can be solved by standard Fourier transforms. 
Considering the thermodynamic limit of $E(L)/L$ and by taking into account
the expression of $\sigma(\lambda)$, we find that the ground state energy per
particle at half-filling is given by the expression
\EQ
e_{\infty} = -4 \sin(\gamma)  \int_{0}^{\infty} d \omega \frac{ 
\ch[\omega(\pi/2-\gamma)] \sh[\omega(\pi-\gamma)]}{\ch[\omega \pi/2] \sh[\omega
\pi]} +\cos(\gamma)
\EN 

We would like to conclude with the following remarks. The experience we have
gained solving the Hamiltonian (1) by the Bethe ansatz is also helpful if one
wants to determine the eigenvalues of the transfer matrix of the underlying
classical statistical model. Exact information about these eigenvalues is
of considerable interest, because the transfer matrix is the generator of the
many conserved currents commuting with the Hamiltonian (1). Considering
that the eigenvalues $\Lambda(\lambda,\{ \lambda_j,\mu_j \})$ of the transfer
matrix are analytical functions of the variables $\{ \lambda_j , \mu_j \}$, it is not difficult to start with an ansatz  fulfilling this requirement. This
is a phenomenological approach which goes by the name of analytical Bethe
ansatz. Taking into account the Bethe ansatz equations (8,9) (analyticity 
conditions), the unitary and the crossing properties of the transfer matrix
, we  can  find the following expression for the  eigenvalues  
\bear
\Lambda(\lambda,\{\lambda_{i}, \mu_j \} ) = 
[a(\lambda)]^{L} \prod_{i=1}^{N_{e}} 
\frac{ \sh(\lambda_i/2-\lambda/2-i\gamma/2)}
{\sh( \lambda_i/2 -\lambda/2 +i \gamma/2)} 
+ [b(\lambda)]^{L} \prod_{i=1}^{N_e} 
\frac{ \ch(\lambda/2-\lambda_i/2-i\gamma/2)}
{\ch( \lambda/2 -\lambda_i/2 + i \gamma/2)} 
\nonumber \\
- [c(\lambda)]^{L} \left \{ 
\prod_{i=1}^{N_{e}} 
\frac{ \sh(\lambda/2-\lambda_i/2+i\gamma/2)}
{\sh( \lambda/2 -\lambda_i/2 -i \gamma/2)} 
\prod_{j=1}^{N_{+}} \frac{ \sh(\mu_j -\lambda +2i \gamma)}{\sh(\mu_j
-\lambda)} + 
 \right. \nonumber \\ \left.
\prod_{i=1}^{N_{e}} 
\frac{ \ch(\lambda/2-\lambda_i/2-i\gamma/2)}
{\ch( \lambda/2 -\lambda_i/2 +i \gamma/2)} 
\prod_{j=1}^{N_{+}} 
\frac{ \sh(\lambda - \mu_j  +2i \gamma)}{\sh(\lambda - \mu_j)}
\right \}
\ear
where the functions $a(\lambda)$, $b(\lambda)$ and $c(\lambda)$ govern the
behaviour of the transfer matrix on the pure ferromagnetic pseudovaccum (
this is equivalent to the totally empty or full band of electrons). For 
instance, if we take into account the $R$-matrix elements of ref. \cite{DEG} 
we find that
\EQ
a(\lambda) = \frac{\sh(i\gamma -\lambda/2)}{\sh(i\gamma+\lambda/2)},~~
b(\lambda) = -\frac{\th(\lambda/2)}{\th(i\gamma+\lambda/2)},~~
c(\lambda) = -\frac{\sh(\lambda/2)}{\sh(i\gamma+\lambda/2)}
\EN
and from equations (15,16) it is clear that all the residues of 
$\Lambda(\lambda,
\{ \lambda_j,\mu_j \})$ vanish, provided the variables $\{ \lambda_j, \mu_j \}$
satisfy the Bethe ansatz equations. We believe that this expression
can  also be derived more rigorously by using our recent 
formulation of the quantum inverse scattering method for Hubbard-like models
\cite{PM}.

In summary, we have solved a supersymmetric model of correlated 
electrons by the Bethe ansatz approach. The ground state energy at
half-filling as well as the corresponding transfer matrix eigenvalues have
been also determined. Preliminary analysis indicates that the excitations
in the model are gapless near half filling. Finally, from the Bethe ansatz
point of view, the limit $\gamma \rightarrow 0$ needs to be performed having
extra care. Due to the peculiar toplogy of roots $\mu_j$ (see equation (11) ),
the situation is analog to the $Su(3)$ isotropic limit of the Izergin-Korepin
model \cite{RE}. We also remark that when both $\lambda_j$ and $\mu_j$ are
real roots, one easily recover the $t-J$ sector present in the $Sl(2|2)$ model
\cite{EKS,EKS1}.

\section*{Acknowledgements}
This  
work was supported by  FOM (Fundamental Onderzoek der Materie) 
and Fapesp ( Funda\c c\~ao
de Amparo \`a Pesquisa do Estado de S. Paulo).

\end{document}